# An analysis of visitors' behavior in the Louvre Museum: A study using Bluetooth data

**Yuji Yoshimura, Stanislav Sobolevsky, Carlo Ratti**
SENSEable City Laboratory, Massachusetts Institute of Technology, 77 Massachusetts Avenue, Cambridge, MA 02139, USA;

**Fabien Girardin**
Near Future Laboratory, Sàrl, CP242, 3960 Sierre, Switzerland;

**Juan Pablo Carrascal, Josep Blat**
Information and Communication Technologies Department, Universitat Pompeu Fabra, Roc Boronat, 138, Tanger Building 08018 Barcelona, Spain;

**Roberta Sinatra**
Center for Complex Network Research and Department of Physics, Northeastern University, 110 Forsyth Street, Boston, MA 02115, USA;

**Abstract.** Museums often suffer from so-called "hyper-congestion", wherein the number of visitors exceeds the capacity of the physical space of the museum. This can potentially deteriorate the quality of visitor's experience disturbed by other visitors' behaviors and presences. Although this situation can be mitigated by managing visitors' flow between spaces, a detailed analysis of the visitor's movement is required to fully realize and apply a proper solution to the problem. This paper analyzes the visitor's sequential movements, the spatial layout, and the relationship between them in large-scale art museums – Louvre Museum – using anonymized data collected through noninvasive Bluetooth sensors. This enables us to unveil some features of visitor's behavior and spatial impact that shed some light on the mechanism of the museum overcrowding. The analysis reveals that the visiting style of short and long stay visitors are not as significantly different as one could expect. Both types of visitors tend to visit a similar number of key locations in the museum while the longer stay type visitors just tend to do so more extensively. In addition, we reveal that some ways of exploring the museum appear frequently for both types of visitors, although long stay type visitors might be expected to diversify much more given the greater time spent in the museum. We suggest that these similarities/dissimilarities make for an uneven distribution of the quantity of visitors in the museum space. The findings increase the understanding of the unknown behaviors of visitors, which is key to improve the museum's environment and visiting experience.

## 1. Mesoscopic research of visitors' sequential movement in Art Museum

Falk and Dierking argue that "a major problem at many museums is crowding, and crowds are not always easy to control" (Falk and Dierking, 1992, p145). Museums and their exhibits along with their own spectacular architecture become one of the most popular destinations for the tourism experience, thus triggering "hyper-congestion" (Krebs et al. 2007), as the number of visitors often exceeds the capacity of spaces, which results in the museum becoming overcrowded.

The congestion in museums shows, on one hand, high attractiveness and vitality, resulting in positive economic impact. On the other hand, the increase of the quantity of visitors implies potential negative effects, which deteriorates the quality of visiting conditions and their experience disturbed by other visitors' behaviors and presences (Maddison & Foster, 2003, page 173-174). In the age when museums play an important role for a massive cultural consumption and urban regeneration with the promotion of the image of cities (Hamnett and Shoval, 2003), they are expected to achieve these seemingly contradictory objectives at the same time; to increase the quantity of visitors and enhance the quality of their experience with achieving the comfortable visit conditions through managing visitors' flow.

Visitors' movement and circulation patterns are recognized as an important topic for museum research (Bitgood, 2006, page 463). However, most of these studies have been conducted at only two extreme cases for art museums: a) visiting patterns on the macro scale to investigate a basic demographic composition of the museum's visitors (Schuster, 1995), along with psychographic factors, which influence visit motives and barriers (Hood, 1983), and b) on the micro scale to research their circulation in the individual exhibition rooms, limited galleries or other areas. This often results in revealing the visitor's attributive features from a socio-cultural point of view (i.e., highly educated and wealthy upper/middle class tend to visit more frequently than the lower social classes) (Hein 1998, page 115-116). Conversely, they have revealed a local interaction between the layout of the exhibit displays of the galleries and visitors' behavior in the exhibition (Melton, 1935; Weiss and Boutourline, 1963; Parsons and Loomis, 1973; Klein, 1993). This polarized research resulted in a shortage of mesoscopic empirical analysis of visitors in large-scale art museums, which has different research targets compared to a single exhibition, the small/medium size museums (Serrel, 1998; Tröndle et al., 2012) or other kind of museums (Laetsch et al., 1980; Sparacino, 2002; Kanda et al., 2007).

Space Syntax (Hillier and Hanson 1984; Hillier, 1996) applies a different approach to analyze the influences of the spatial layout and design of buildings from the formation of visitors' movement and behavior by describing the overall configuration of the museum setting (see Hillier and Tzortzi 2006 for a review). This type of knowledge is key to produce patterns of exploration and interaction of visitors, and the co-presence and co-awareness between visitors in the museum environments as a whole (Choi, 1999).

Yet all of these studies rely on a spatially and temporally limited dataset, which often results in providing just a snapshot of a limited area in the built environment. Even a simulation-based analysis requires a simplification of the human behavior, to estimate visitors' behavior rather than reveal actual patterns of movement with real world empirical data.

This paper analyzes visitor's sequential movement, the spatial layout, and the relationship between them in order to clarify the behavioral features in the large-scale art museum – Louvre Museum. We focus on visitors' circulation from the entrance to exit as a whole mobility network rather than movement in particular individual rooms. The way of visiting exhibits is analyzed by means of the length of stay and visiting sequence order, because these visit conditions determine the visitor's perceptions and attentions thus shaping the visitor's experience (Bitgood, 2006). The length of stay might be thought to be the key factor determining the number of visited places and the order of visiting them, resulting in a variety of different kinds of the routes; the more time you are given, the more opportunity you have, and vice versa. The question to be asked is whether this hypothesis is actually true, and by its extension, how the length of stay and the order of visited places make visitors' mobility style different, and how this dissimilarity is seen in the museum. This understanding might be the key to improve the museum environment as well as to enhance visitors' experience.

We employ the systematic observation method relying on Bluetooth proximity detection, which makes it possible to produce the large-scale datasets representing visitors' sequential movement with low spatial resolution. "Large-scale datasets" refers to that the sample size of this paper is much larger than the previous studies' ones collected in art museums (i.e., almost 2,000 (Melton, 1935); 689 (Serrell, 1998); 576 (Tröndle et al., 2012); 50 (Sparacino, 2002)), although each of them contains different kinds of information with a high enough resolution for their particular objectives, as human-based observation, GPS, RFID or ultra-wideband technology can achieve. In the present work we explore the global patterns of visitors' behaviors by increasing the quantity of data, because "when we increase the scale of the data that we work with, we can do new things that weren't possible when we just worked with smaller amounts" (Mayer-Schönberger & Cukier, 2013, p10).

Thus, our research limits to deal with visitors' physical presence in and between places without questioning their introspective factors (i.e., learning process, making meaning from the experience of the museum), which the previous studies tried to answer by small scale sample (see Kirchberg & Tröndle, 2012 for a review). Conversely, the superimposition of huge amounts of individuals' behaviors and their changes in time makes some patterns appear to be self-organizing in a bottom up way from seemingly chaotic, disordered and crowded movement. These results could shed light on the quality of the visit conditions derived from the overcrowding, not only around the spots where the iconic art works are placed, but also the spaces along them with the dynamic visitors' flow in the network. The better understanding of visiting features helps in designing more adequate spatial arrangements and gives insights to practitioners in order to manage visitors' flow in a more efficient and dynamic way.

## 2. Visitor's sequential movement and analysis framework

The use of large-scale datasets enables us to discover and analyze the frequent patterns between human activities. These analyses have been conducted in the specific spatiotemporal limitations derived from the limited measurement of mobile objects (Miller, 2005) at different contexts and at various scales in order to shed light on unknown aspects of the human behaviors: to discover the patterns of human mobility (González et al., 2008) and the urban activities (Ratti et al., 2006) through cell phone usage at the regional scale, to analyze the sequential patterns of tourists at the local scale by the number of visited locations, its order and the length of stay by GPS data (Shoval et al. 2013), and for instance to disclose some aspects of customers' purchasing behavior in the grocery store by analyzing customer's path, their length of stay and the categories of the purchased products through RFID data (Hui et al. 2009).

Our previous research proposed a Bluetooth based data collection technique in a large-scale art museum at the mesoscopic scale in order to classify visitors' behavior by their most used paths and their relationship with the length of stay (Yoshimura et al. 2012). Bluetooth detection is based on the systematic observation, which discover Bluetooth activated mobile devices, in the framework of the "unobtrusive measures", making use of unconsciously left visitors' *digital footprint*. The considerable number of researches has employed this method but not in the context of the large-scale art museums: for measuring the relationship of the social network between people (Eagle and Pentland, 2005; Paulos and Goodman, 2004), for analyzing mobility of pedestrian and their relationships (Kostakos et al., 2010; Versichele et al., 2012; Delafontaine et al., 2012), and for estimating travel times (Barceló et al., 2010).

A Bluetooth proximity detection approach to the analysis of visitor's behavior in museums has many advantages. Contrary to the granular mobile phone tracking (Ratti et al, 2006), the detecting scale by Bluetooth is much more fine grained. In addition, in contrast to RFID tags (Kanda et al. 2007; Hui et al., 2009) and active mobile phone tracking with or without GPS (Asakura and Iryob, 2007), previous registration is neither required nor necessary to equip any devices or tags in advance with Bluetooth. The fact that no prior participation or registration is required by visitors enables a mass participation of subjects to collect an enormous amount of data in the long term, contrary to the time constraint case (McKercher et al. 2012; Shoval et al. 2013). Also, the unobtrusive feature of Bluetooth removes bias in data, which could be created from a visitor consciously being tracked. Furthermore, Bluetooth proximity detection succeeds inside buildings or in the proximity of tall structures, where GPS connectivity is limited. All of those advantages make this method adequate for generating visitors' sequential movement between key places without specifying their activities, attributes nor inner thoughts in a consistent way at the mesoscopic scale in the large-scale art museum.

We identify visitor's length of stay at a certain location as the indicator of measuring their interest level at that exhibit, by merely accounting for their presence without questioning their inner thoughts. Also, we estimated visitors' routes between sensors and its quantity at the place by the collected data.

As all of our analysis and the interpretation of data are conducted within the specific spatio-temporal framework, therefore our approach has some limitations. Firstly the concept of trajectory used in this paper is different from the one usually available when working with data collected by GPS systems. This is because a Bluetooth proximity sensor just let us know the time-stamped sequence of individual transitions between nodes (e.g. sequence of A-B-D) of a mobile device, while a GPS system can track all the movements of a device. However the network of rooms derived from the spatial layout of the museum determines the feasible routes, and this enhances estimation of the used paths by visitors between sensors without observing their exact trajectories and loads per room (Delafontaine et al., 2012). Secondly, we can't deal directly with visitors' introspective factors, which are visitors' expectations, experiences and satisfactions (Pekarik, et al. 1999). This results in excluding "wayfinding", which refers to visitors' ability to find his/her way within a setting, and "orientation", which indicates an available knowledge in a setting, with the use of the hand-held maps and direction signs, as research questions from our study. Because they consist of the complex interaction between the environmental cognition and those orientation devices. In addition, visitors' presence at the specific place is not necessarily related to their engaging time for the exhibits, while previous studies used it for measuring visitors' interest (Robinson, 1928; Melton, 1935). Finally, our sample has a possible bias in two ways: the sample composition is affected by the segments of the mobile device holders and their decision to activate/not activate Bluetooth function. Although the latter requires calculating the sample representativeness and is typically conducted by using a short term manual counting (Versichele et al. 2012), we employed the long-term systematic comparison during one month; the number of devices detected at the entrance with the official museum head counts and ticket sales. This method provided us with more comprehensive information compared to the previous researches.

## 3. Concept Definitions and Data Settings

This section defines the locations of sensors used and the components of the dataset in order to explore our method and data consistency. We collected a dataset during a specific period and processed it into an specific form required for analysis.

### 3-1. Sensors settings in museum and definition of node

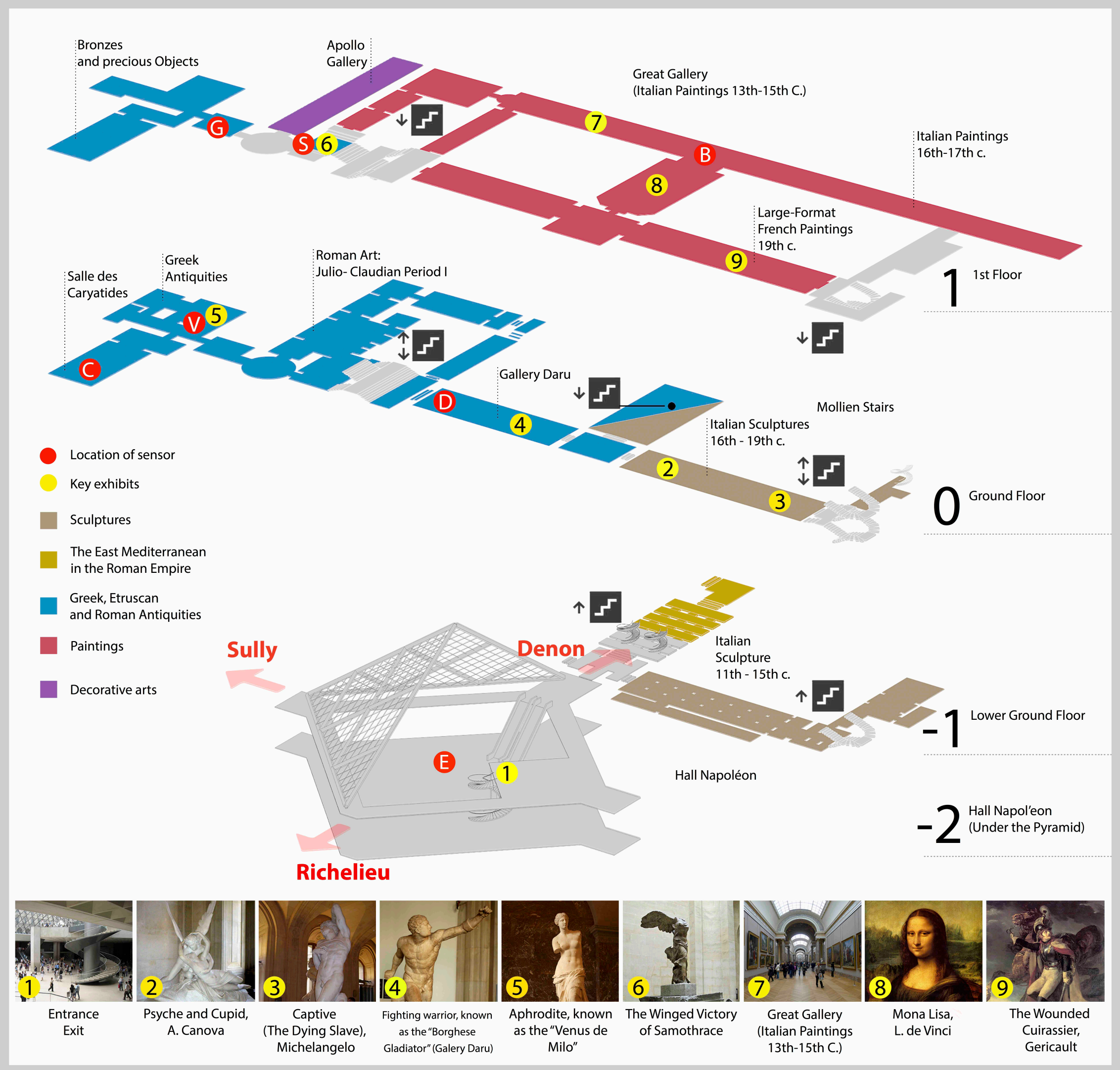

**Figure 1.** [In colour online] Location of 7 sensors indicating their approximate sensing range

The Figure 1 shows the location of 7 sensors deployed all over the museum covering key places to detect visitors. They are situated in one of the busiest trails identified by Louvre Museum, which lead visitors from the entrance to the Venus de Milo; Hall (E),

Gallery Daru (D), Venus de Milo (V), Salle des Caryatides (C), Great Gallery (B), Victory of Samothrace (S) and Salle des Verres (G).

Each sensor forms a detectable area, which is identified as a *node*, approximately 20 meters long and 7 meters wide. This area fluctuates depending on various museum settings, including the location of sensor (e.g. inside functional wooden boxes, desks, or open space). However, all sensors covered targeted areas along the paths to key museum iconic art works. Once a Bluetooth activated mobile device enters the detectable area, the sensor continues to receive the emitted signal from the mobile device until it disappears. Thus, the sensor registers the time at which signal of a mobile device appears, also called check-in time. Afterwards, when the signal of a mobile device disappears, the sensor records the check-out time. Then the time difference between each mobile device's check-in and check-out time can be calculated. This defines the length of stay at the node. Similarly, by looking at the first check-in time and the last check-out time over all nodes provided that the first and last nodes correspond to entry point and exit from the museum, it is possible to calculate how long a visitor stays in the museum. The series of check-in and check-out time data registered by all the installed sensors makes it possible to construct the visitor's trajectory in the museum. In addition to the length of the stay, the sensors timestamp the data allowing calculation of the travel time between nodes. The synchronization of all sensors makes it possible to perform fine-grained time series analysis. All of this information can be achieved without invading visitor privacy, because the SHA algorithm (Stallings, 2011, page 342-361) is applied to each sensor where the MACID is converted to a unique identifier (Sanfeliu et al, 2010).

### 3-3. Collected Sample

We collected data over 24 days; from 30/April to 9/May 2010, 30/June to 8/July 2010, and 7/August to 18/August 2010. We selected data, starting and finishing at node E in order to measure the length of stay in the museum. Consequently, 24,452 unique devices were chosen to be analyzed for this paper. On average, 8.2% of visitors activated Bluetooth on their mobile device in the Louvre Museum (Yoshimura et al, 2012).

**Data cleanup:** The data collection was performed at different periods by a different number of sensors. We checked for possible synchronization issues due to lack of calibration, then adjusted the data to remove any inconsistencies. Finally, we only used data from visitors who started from node E and finished at node E to be able to measure complete length of stay in the museum – entries like this indicate that the visitor was correctly registered when he/she entered, moved inside, and left the museum.

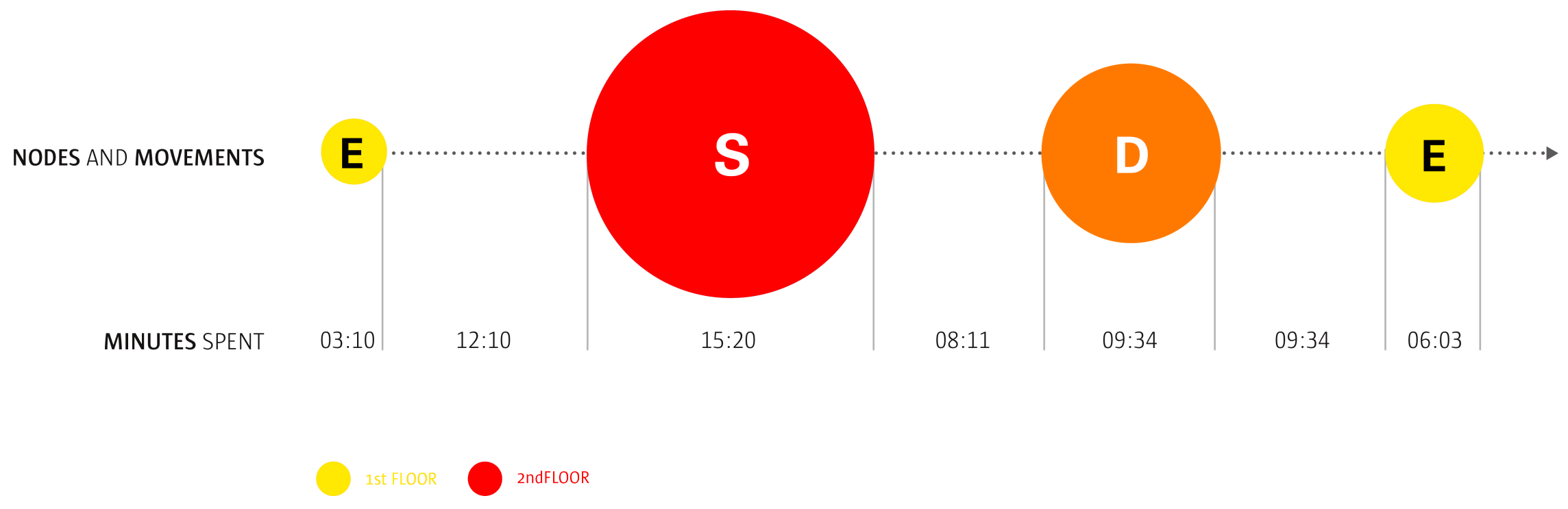

**Figure 2.** [In colour online] Visualization of a relationship between the sequential movement and the time of stay for a visitor

**Data processing:** Figure 2 graphically shows the features of the logged data. It displays all entries of a visitor in the database for one day. Every circle with an alphabetic letter symbolizes detection at a certain node. It indicates that this visitor made a sequential movement, E-S-D-E, and stayed at node E for 3 min 10 seconds, node S for 15 min 20 seconds, node D for 9 min 34 seconds and, again, node E for 6 min 3 seconds. The travel times between corresponding nodes are: 12 min 23 seconds for E-S, 8 min 11 seconds for S-D and 9 min 34 seconds for D-E.

**Table 1.** Example of the dataset

| Rffr | Date | Path | checkin | checkout | staylength |
|---|---|---|---|---|---|
| Unique ID | 2010-04-30 | E-S-D-E | 09:04:35 | 11:07:52 | 02:03:17 |

We build a database and designed a query engine to extract and transform the data for the different stages of the analysis. Table 1 shows an example of components of the transformed dataset. There is one entry per visitor, and it includes the date of the visit, the path followed across the museum, the time of entrance to the museum (check-in), time of exit (check-out), and the total length of stay inside the museum.

### 3-4. Partitioning of Visitors

In order to find the characteristics, the typical patterns of visits and other determinant features of visitors' behavior, we examined two extreme groups. Firstly, we sort all the visits of our sample size (24,452 visits) by their total time spent in the museum. By binning them into deciles, we obtain equally-sized clusters of ~2446 visits each. We refer to all the visits, which are found in the first decile as "short visits" and respectively refer to these visitors as "short stay type visitors". Similarly, we refer to the visits of the tenth decile as "long visits" and these visitors as "long stay type visitors".

## 4. Results

In the following subsections, we present an overview of the statistical analysis built around the previously described dataset. We discuss the path sequence length, which is the number of visited nodes including the multiple visits without "E", the length of the visitors' path, and the frequency of the appearance of each path. The distribution of the path sequence length is also presented and analyzed. We reveal visiting patterns, and the similarity and dissimilarity of the behaviors of the longer and shorter stay type visitors.

### 4-1. Basic statistics of visitors' behavior

We analyzed all visitors' data to capture the features of their behavior focusing on the path sequence length and its relationship with the length of stay in the museum.

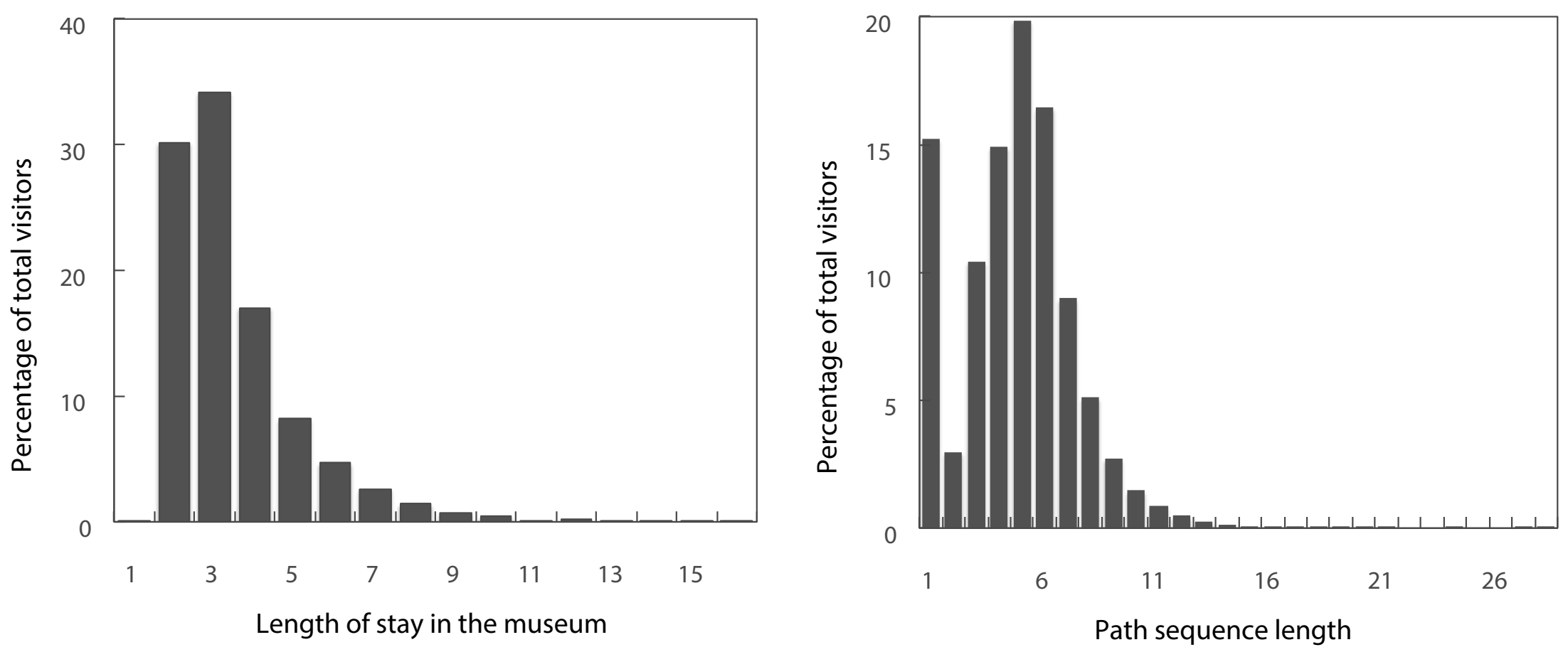

(a) (b)

**Figure 3.** [In colour online] (a) The distribution of visits per the length of stay in the museum. (b) The distribution of the path sequence length.

Figure 3 (a) shows the distribution of the number of visits (y-axis) per the length of stay in the museum binned for each hour (x-axis). Although the maximum length of stay is more than 15 hours, only 410 visitors stayed for more than 8 hours, which corresponds to 1.6% of the total. Conversely, the minimum length of stay of less than 1 hour happens for only one visitor, while more than 30% of visitors stayed for 1-2 hours. Those facts indicate that the extreme visitors, whose length of stay is more than 8 hours or less than 1 hour, can be aggregated for the statistical reliability without substantially affecting the time-sensitive behavioral analysis. The distribution of the length of stay is positively skewed, with the majority of the visitors having the length of the stay between 4-6 hours.

Next, we look at the distribution of the path sequence length (see Figure 3. (b)). Although the maximum length of the path sequence length is 30, the percent of visitors who visited more than 15 nodes is only 0.5%. In general, this plot shows a distribution slightly skewed to the right, but visitors who visited only 1 node appear quite frequently, covering 15.2% of the total. Visitors who visited 2 nodes rarely appear (i.e., 2.9%). However, the length of the sequence by itself doesn't necessary reveal the size of the visitor mobility area, because a visitor could easily move between the nearby nodes frequently without radially expanding throughout the museum.

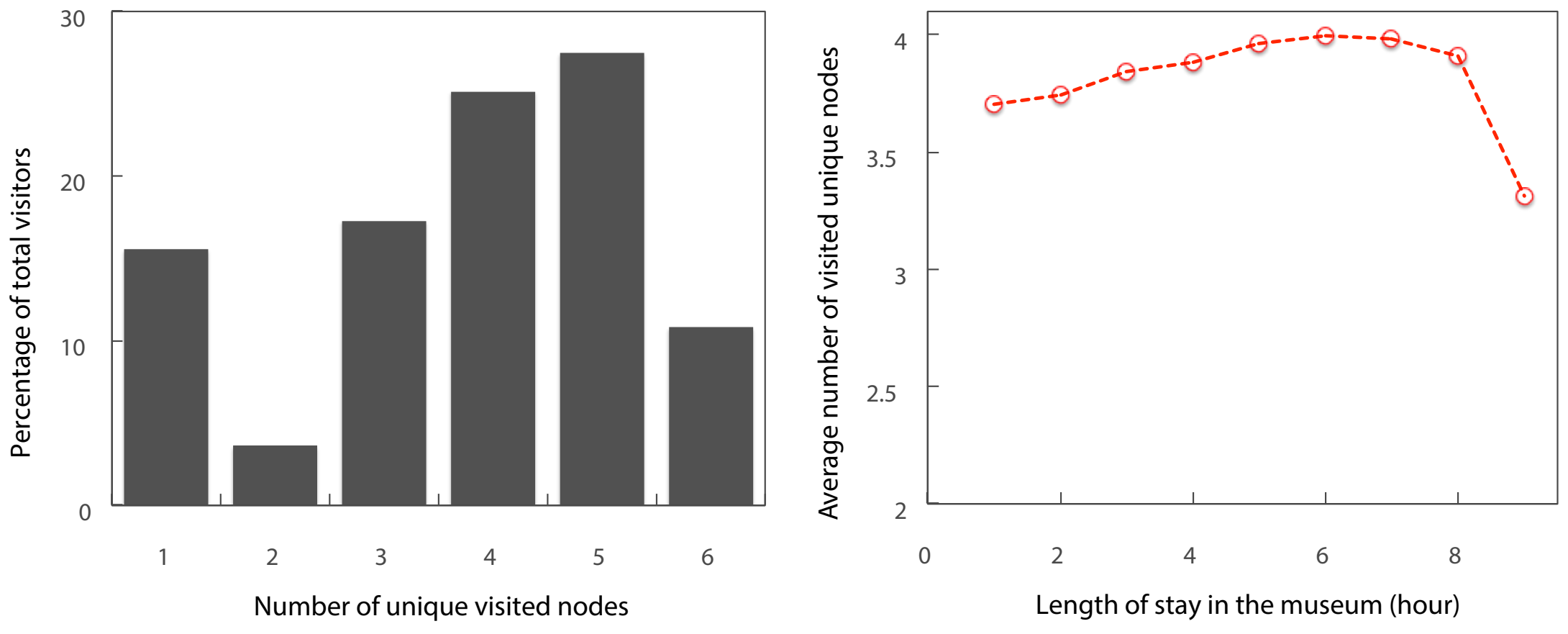

(a) (b)

**Figure 4.** [In colour online] (a) Distribution of the number of unique nodes visited other than "E". (b) The average number of visited unique nodes per the duration of the visit.

Figure 4 (a) represents the number of unique nodes visitors passed during their stay in the museum. We can observe that visiting 2 nodes rarely happened, while visiting 1 and 3 nodes almost have the same frequency. The most frequent number of unique nodes visited is 4 or 5 nodes, while visiting all 6 nodes rarely happens as well. This indicates that some factors in most of the cases prevent from exploring all the nodes, while all nodes but one already can be explored much more often. In addition, Figure 4 (b) reveals that the average number of visited unique nodes per the duration of the visit is almost constant. The correlation coefficient between these two variables (Spearman's correlation=0.072, p-value<2.2e-16) indicates that the unique number of visited nodes is independent from the duration of stay in the museum, and vice versa. Surprisingly the longest stay visitors use to visit even less nodes on average compared to the shorter-stay ones.

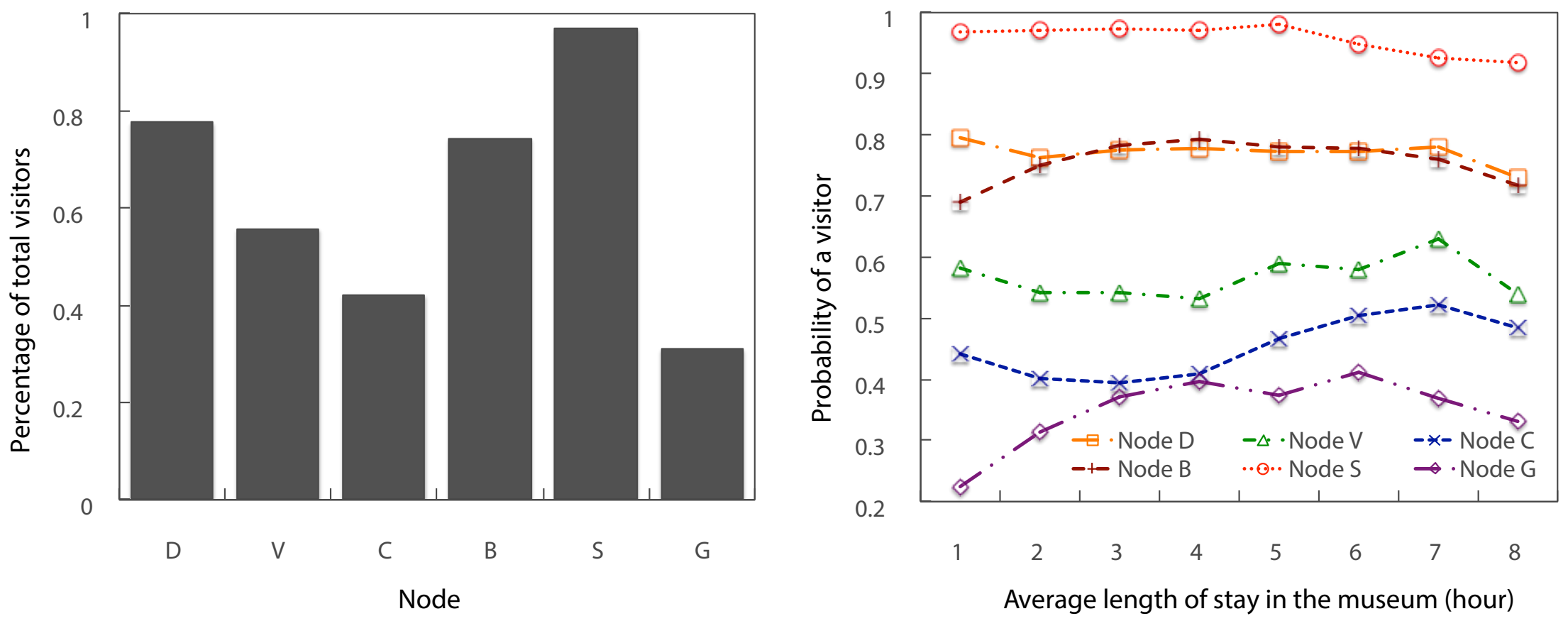

(a) (b)

**Figure 5.** [In colour online] (a) The frequency of visits each node receives. (b) The frequency of visiting different nodes at least once against the duration of stay.

Figure 5 (a) shows the frequency of visits for each node. 97% of all of visitors passed node S. Nodes D and B are frequently visited (i.e., nearly 80% of both cases). On the other hand, node G is the most rarely visited (i.e., just 30% of all of visitors). Figure 5 (b) presents the attractivity of the nodes depending on the duration of the visit. As we can see, the probability of visiting most nodes doesn't depend on the length of stay in the museum, as this trend is nearly constant for all nodes. Within them, node G behaves differently from the others, as its probability increases with the visitors' length of stay in the museum. This shows that shorter stay type visitors show a lower tendency to visit node G, while longer stay type visitors seem more attracted to visit this node (perhaps having more time to explore this part of the museum), although its frequency doesn't surpass 40%, regardless of the visitor type.

**Table 2.** Two types of visitors' transition rate from previous nodes to node G expressed as a percentage.

| Current location/node G | Shorter stay type | Longer stay type | Their difference |
|---|---|---|---|
| D | 4.00% | 7.17% | **3.17%** |
| V | 1.38% | 3.17% | 1.79% |
| C | 4.86% | 9.91% | **5.05%** |
| B | 5.60% | 6.30% | 0.70% |
| S | 2.53% | 5.69% | **3.16%** |

We can observe this change on the transition rate (probability of moving to the given destination node right after visiting the given origin) from any other node to node G on two types of visitors (see Table 2). All of the transition rates increase as the visitor's length of stay increases. Within them, node D, node C and node S show substantial increases in bold as seen in Table 2.

### 4.2. Similarity of visitors' behaviors

By looking at the path length of the visitors of different stay time we find another surprising effect – though the path length slightly increases with increase of stay time, the path length of long stay type visitors is not that substantially longer than of the short stay visitors, as one may expect. In addition, the number of nodes that makeup one's stay are very similar.

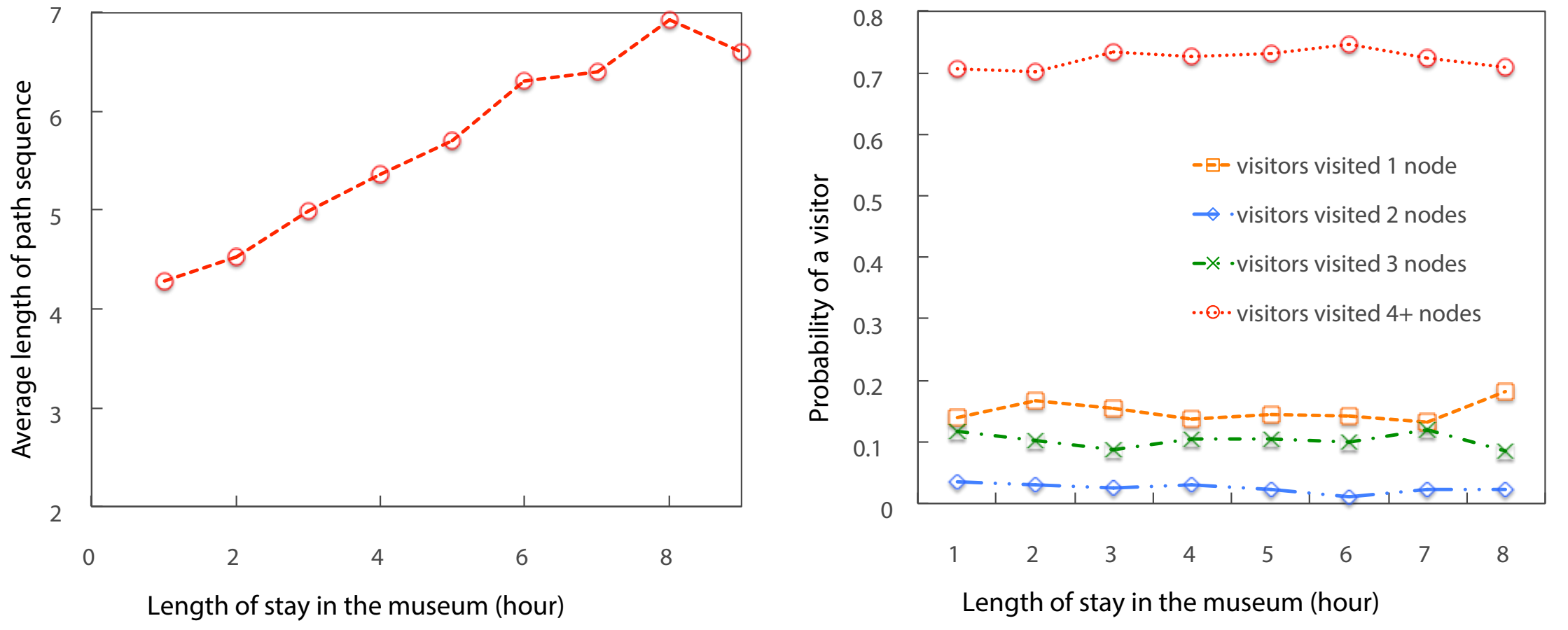

(a) (b)

**Figure 6.** [In colour online] (a) The average length of path sequence (y-axis) against the average length of stay in the museum (x-axis). (b) The probability of a visitor's path length being 1, 2, 3 or more nodes by their length of stay in the museum (x-axis)

**Table 3.** The average length of path sequence per each hour and its percentage of increase

| Hours | Path sequence length | Percentage of increase |
|---|---|---|
| 1-2 | 4.28 | 5.84% |
| 2-3 | 4.53 | 9.93% |
| 3-4 | 4.98 | 7.63% |
| 4-5 | 5.36 | 6.34% |
| 5-6 | 5.70 | 10.53% |
| 6-7 | 6.30 | 1.43% |
| 7-8 | 6.39 | 8.29% |

| 8-9 | 6.92 | - 4.62% |
|---|---|---|

**Table 4.** The probability of a visitor to have a path length of 1, 2, 3 or more by the length of their stay in the museum.

| *Length of stay (hour)* | *Visitors visited 1 node* | *2 nodes* | *3 nodes* | *More nodes* |
|---|---|---|---|---|
| 1-2 | 0.14 | 0.03 | 0.11 | 0.70 |
| 2-3 | 0.16 | 0.03 | 0.10 | 0.70 |
| 3-4 | 0.15 | 0.02 | 0.08 | 0.73 |
| 4-5 | 0.13 | 0.03 | 0.10 | 0.72 |
| 5-6 | 0.14 | 0.02 | 0.10 | 0.73 |
| 6-7 | 0.14 | 0.01 | 0.10 | 0.74 |
| 7-8 | 0.13 | 0.02 | 0.12 | 0.72 |
| 8-9 | 0.18 | 0.02 | 0.08 | 0.70 |
| Average | 0.14 | 0.02 | 0.10 | 0.72 |
| Standard Deviation | 0.01 | 0.00 | 0.01 | 0.01 |

Figure 6 (a) reveals that, while visitors tend to visit on average 4.3 nodes when staying in the museum for 1-2 hours, they are likely to visit only 5.5 nodes when they stay for 3-7 hours. The latter's length of stay is 3 times longer than the preceding, while it results in only 28% of the sequence length increase. In addition, the longer stay type visitors (i.e., 9-10 hours) show that they visited 6.6 nodes on average, which is even less than that of the 8-9 hour visitors. The path sequence length increases as the duration of stay increases, but the rate of change is not that substantial (see table 3) especially if compared to durations of increase. Figure 6 (b) presents the probability for visitors to have a certain path length versus their length of stay in the museum. The probability to visit 1, 2, 3 or more nodes, against the length of stay aggregated by each hour, appears almost flat, suggesting its independence on duration of stay in the museum. We can also observe this tendency by examining the frequently appearing paths from the shorter and longer stay type visitors.

**Table 5.** Top 5 of the frequently appearing paths from the longer stay type and shorter stay type of all visitors.

| Longer stay type visitor; its frequency | Shorter stay type visitor; its frequency |
|---|---|
| *Visitors whose length of path is more than 4* | |
| E-D-S-B-D-V-C-E; 2.23% | E-D-S-B-D-V-C-E; 8.26% |
| E-D-S-B-D-E; 1.84% | E-D-S-B-D-E; 6.89% |
| E-D-S-B-D-V-E; 1.50% | E-D-S-B-V-E; 5.57% |
| E-D-S-B-D-S-E; 0.89% | E-S-D-V-C-E; 4.67% |
| E-D-S-B-D-G-E; 0.78 | E-C-V-D-S-B-E; 3.71% |
| *Visitors whose length of path is less than 4* | |
| E-S-E; 45.10% | E-S-E; 36.34% |
| E-D-S-B-E; 12.38% | E-D-S-B-E; 16.62% |
| E-S-E-S-E; 7.64% | E-V-D-S-E; 4.51% |
| E-B-E; 5.19% | E-G-S-B-E; 4.12% |
| E-G-S-B-E; 4.28 | E-S-B-E; 3.35% |

Table 5 presents the top 5 most frequently appearing paths of the short and long stay type visitors. We counted the number of paths, which appear in both groups and divided by the total number of visitors in each group (i.e., 2,445), in order to obtain the frequency of a path appearing. This reveals that both groups have similar frequent path length; i.e., the short stay type paths are just slightly shorter compared to the long stay paths. The first and second frequently appearing paths from the long and short stay type in both groups are very similar, otherwise the frequency of the group that visited more than 4 nodes is much lower than those who visited less than 4 nodes. The results show that the behavioral ways of short and long stay visitors are not as significantly different as one could expect. Both types of visitors tend to visit the same number of popular places while the longer stay visitors just tend to do so more extensively.

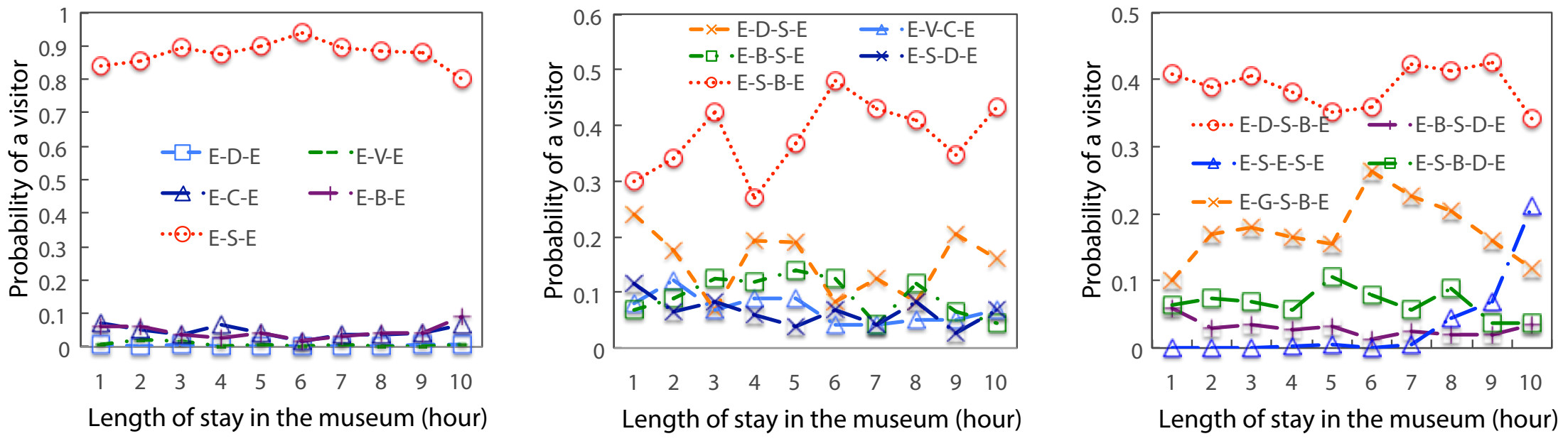

(a) (b) (c)

**Figure 7.** [In colour online] (a) The probability for visitors to make particular paths of length 1. (b) The probability for visitors to make particular paths of length 2. (c) The probability for visitors to make particular paths of length 3 versus their length of stay in the museum.

Let's examine in more detail the visitors whose path length is less than 4 nodes. Within them, the most frequently appearing path for each category (i.e., visited 1 node, 2 nodes, 3 nodes) coincides well between both groups of shorter and longer stay type visitors. Figure 7 presents the probability of visitors who visited 1 node (a), 2 nodes (b) and 3 nodes (c). We can observe that only one path per each group has a strong influence on the probability as a whole especially in Figure 8 (a) (i.e., 89.4% of those visitors made the path E-S-E).

Similarly, visitors who follow the E-S-B-E path, which is the most frequently appearing path in the group that visited 2 nodes (i.e., 37.52%), and E-D-S-B-E, which is the most frequently appearing path in the group that visited 3 nodes (i.e., 38.94%), just visit one more node at the end of their visit. There is no clear difference between the long and short stay type visitors in those three groups; rather, their behaviors seem very similar, other than the substantial difference in the length of stay.

## 6. Discussion

The previous sections revealed that many features of the longer and shorter stay type visitor's behavior including the path sequence length as well as the visited unique nodes, do not appear to be strikingly different among visitors of different duration of stay, and sometimes even independent or nearly independent of it. This section uncovers that visitor's path and their variations are quite selective, which visitors mostly choose the same paths although many other options exist, in terms of the path sequence length and the sequential order, thus creating an uneven distribution of visitors among spaces, and possibly one of the largest causes of high congestion and vacancy in the museum.

### 6.1. Uneven spatial distribution of visitors

The interplay between sensor locations and the spatial layout of the museum determines the specific and possible route(s) used by visitors. All sensors were logistically placed in the determinant positions for visitors' route choice in the museum. Therefore, the transition between two places makes it possible to estimate the determinant route where visitors take. Thus, we can clarify the uneven spatial use of the museum accesses for visitors' entry and exit behaviors through analyzing the first or last 2 locations in their sequence: 71.6% of visitors took E-D, E-B, E-S, meaning that they entered through *Denon access*. This indicates that only 28.3% used *Sully* or *Richelieu access* (i.e., E-V, E-C, E-G), while 57.3% of them exited from *Denon access*, suggesting a decrease of 14.3% of the users when they exit. This technique enables us to determine the visited rooms without observing their exact trajectories. Also, this indicates that we could speculate the volume of visitors and their concentration along the specific paths without knowing the exact load per room.

The previous section revealed that 13.5% of total visitors, who visited the museum, just visited the Victory of Samothrace (node S – one of the most iconic exhibits in the museum) without visiting any of the other 5 nodes in the museum. Considering the Museum's spatial layout, these visitors used the *Mollien* stairs, which connect the $16^{th}$ – $19^{th}$ c. Italian sculpture rooms on the ground floor and the $19^{th}$ c. French painting room on the first floor to visit node S, instead of using the Victory of Samothrace staircase where node D is located (see the red line at Figure 8).

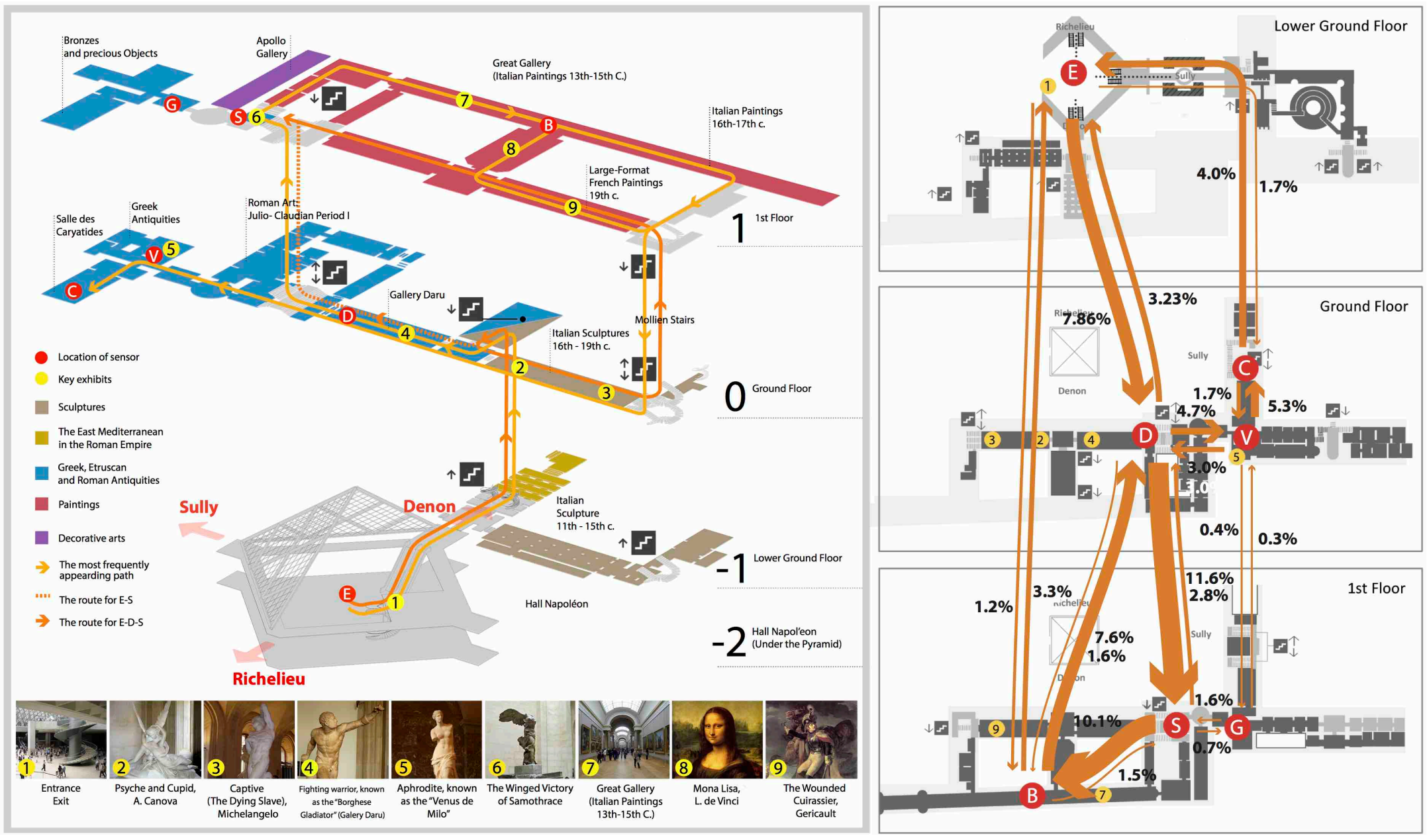

**Figure 8.** [In colour online] (a) The map of the spatial layout of the Louvre museum and the used visitors' routes. (b) The transition percentage between locations, which show only major links between each pair of nodes.

From the spatial point of view, this is the intriguing result because the shortest path from the entrance (i.e., node E) to the Victory of Samothrace (i.e., node S) is the one, which passes by the node D, meaning that they turned to the left at the point Z (see the red dot line at Figure 8). To use the route through the *Mollien* stairs signifies the detour, both spatially and temporally, to reach the node S.

**Table 6.** Three types of visitors' transition rate from node E to the subsequent node expressed by the percentage.

| Subsequent node from node E | All datasets | Shorter stay type | Longer stay type |
|---|---|---|---|
| **D** | **43.32%** | **42.41%** | **40.34%** |
| V | 11.25% | 12.80% | 11.38% |
| C | 9.59% | 10.02% | 11.32% |
| B | 6.80% | 9.20% | 7.51% |
| **S** | **21.53%** | **21.02%** | **20.82%** |
| G | 7.51% | 4.54% | 8.63% |

Table 6 reveals visitors' route choice more in detail; almost 40% of visitors turned to the left to reach node D (i.e., E-D), while around 20% of visitors turned to the right (i.e., E-S). And again, there is no significant difference of the behaviors between the longer and shorter stay type visitors, meaning that both start their museum experience in a similar way. In addition, all of those visitors, who made E-S-E, tend to stay in a very confined area of the *Denon* Wing during their visit, because node D, node V, node B and node G are installed in some key points of the exit from *Denon* Wing. They just explore and stay in the small area during their visit, and this tendency can be found even more on the longer stay type visitors than the shorter stay ones (see Table 5).

On the other hand, the most frequently appearing path of both groups that visited at least 4 nodes is E-D-S-B-D-V-C-E; where the visitor visited the Gallery Daru, the Victory of Samothrace, the Great Gallery and the Venus de Milo (see the blue line at Figure 8). This path starts from a trail of E-D and finish with C-E, indicating that the visitor entered the museum from the *Denon* access, and exited from the *Richelieu* or *Sully* access. This suggests that visitors tend to explore the extensive places of the museum through covering most of the iconic exhibits rather than staying in only a part of the museum. In addition, the frequency of this path in the shorter stay type visitor is much higher than that of the longer stay type visitors. This could indicate that the shorter stay type visitors might tend to select the spatially most optimized paths to visit all of possible iconic exhibits within their limited available staying time in the museum.

We believe that shorter stay type visitors explore fewer of the popular places due to the limited time that they have to spend in the museum. This is intuitive since a visitor's movement and their activities would be limited when their length of stay time in the museum is short. Consequently, the trajectories of the longer stay type visitor are expected to be more complex than those of the shorter stay type visitors, and vice versa. However, the results show that the behavioral patterns of short and long stay visitors are not as significantly different as one could expect. Both types of visitors tend to visit the similar number of the popular rooms while the longer stay visitors just tend to do so more extensively.

All of them imply that visitors' trajectories seem to be quite limited in terms of the path sequence length and its order, although there exist a number of possible routes including the repeats of the same nodes. More generally, we might say that- and this is partially coincided with Choi (1999)'s statement-, the more the number of spaces to be able to visit increases, the more the visitor's path tends to be selective. That is, when the number of the rooms with exhibits increases, visitors seem not to visit all of the exhibits, but only a few of them selectively. But our findings tells further; these limited paths and their use are almost independent from the length of stay, meaning that the most of visitors, despite of their length of stay is short or long, tend to use the same trajectories.

We speculate that this similarity/dissimilarity of the patterns make the uneven distribution of the quantity of visitors in the museum' space; for instance, the trail of E-D-S-B-D is frequently observed, independently from their length of stay, suggesting that there can be the high concentration of visitors in those enclosed areas. On the contrary, some spaces can be found quite vacant; the sequential pattern between node S and node G is rarely found, especially, in the shorter stay type. This indicates that the topological proximity and the attractivity of the node can be changed depending on the visitor's length of stay (see Figure 5). This could be because node G, which tends to be visited when people have more time to explore the museum, is not seen as a necessary or "priority" during the museum visit. Thus, the distribution of visitors and the number of visits that each room receives becomes unequal.

## 7. Conclusion

This paper examines visitors' mobility styles and their respective spatial impacts by analyzing large-scale datasets obtained through Bluetooth proximity detection in a bottom-up methodology. This analysis and obtained results give a great scientific advancement to improve visiting conditions, which are strongly affected by the quality of a visitor's experience in the museum.

The results indicate that the behavioral ways of short and long stay type visitors are not as different as one might expect. The path lengths grow at a much slower rate compared to the duration of stay. Even more surprisingly, the number of unique nodes visited stays almost constant, independent on the length of stay. The correlation coefficient between these two variables quantitatively indicates that the unique number of visited nodes is independent from the duration of stay in the museum, and vice versa. Both – short and long-time – groups visit mostly the same number of sensor locations, while the longer stay type visitors just tend to do so more time extensively. Moreover, the probability of the appearance of visitors whose path sequence length is small (<4 nodes), is constant across all time divisions, meaning that there always exist a certain category of visitors who don't try to extensively explore museum space no matter how much time do they have to do so. Also we discovered that the frequency of the node visits per hour is almost constant independently of the length of stay in the museum.

Conversely, we can point out key differences in visitors' behavior within each of two groups – those who visited more or less than 4 nodes. The average number of locations visited for each of the groups again does not depend on the time people have to spend in the museum (i.e., independent of a visitor being classified as a shorter and longer stay type visitor). For both – short and long-term visitors – the most frequently appearing path in the group that visited at least 4 nodes is E-D-S-B-D-V-C-E. We might indicate that this path could be one of the most optimized paths, which would enable visitors to explore all of the interesting places as fast as possible. On the contrary, the group that visited just a few nodes (less than 4) which appears to be of relatively the same size among both – short and long stay types of visitors - might be interested in just a few of the iconic art works, or just not motivated or informed enough in order to explore bigger space.

All of those facts suggest that some ways of exploring museum appear as frequent ones for both – short and long time visitors – even though the least might be expected to be much more diverse in their choices given more time available. They imply that visitors' sequential movement in the Louvre Museum is quite limited in terms of the path sequence length and order, though there is the number of possible routes including repeats of the same nodes. We speculate that these similarities/dissimilarities could cause uneven distribution of the quantity of visitors, resulting in the congestion/vacancies in the museum spaces.

These findings present a significant advancement in describing patterns in visitors' activity and behavior in the Museum, and might enable us to foresee the visitor movement. This also indicates the possibility of dynamically managing visitors' flow and museum congestion, taking into account time-related factors, and the possible advantages of designing spatial arrangement. In addition, the transition rate and the probability of movement between nodes makes it possible to foresee the specific quantity and flow of visitors at a certain time and space, helping the development of more flexible and dynamic policies for space control. For instance, the similarities/dissimilarities of both types of visitors, which were unknown prior to this study, might make the practitioner reconsider the target of some management techniques, which should be carefully applied on the proper and segmented group (Krebs, et al., 2007; Maddison & Foster, 2003). Also, a dynamic visitor control system might be developed, based on our findings, by using the audio guides to change visitors' suggested routes dynamically depending on the congestion level as calculated by the data gathered from sensors installed throughout the museum.

Finally, these results might allow improving the quality of information that can be provided to visitors at an adequate place and time in order to maximize their fulfillment of the social and cultural experience thereby optimizing the museum infrastructure.